\documentclass[reprintb,aps,twocolumn,nofootinbib]{revtex4-2}

\newcommand{\be}{\begin{equation}}
\newcommand{\ee}{\end{equation}}
\newcommand{\bea}{\begin{eqnarray}}
\newcommand{\eea}{\end{eqnarray}}

\usepackage{amsmath}
\usepackage{amssymb}
\usepackage{xcolor} 

\begin{document}

\title{Exact solution of two-dimensional Palatini Gauss-Bonnet theory on a strip}

\author{M\'aximo Ba\~nados$^1$ and Marc Henneaux$^2$  \\
{\footnotesize $^1$ Facultad de F\'isica, Pontificia Universidad Cat\'olica de Chile,  Santiago, Chile}  and   
{\footnotesize $^{ 2}$ Universit\'e Libre de Bruxelles and International Solvay Institutes,
Brussels, Belgium} \\ {\footnotesize and Coll\`ege de France,  Universit\'e PSL,   Paris, France\\
Email: mbanados@uc.cl, marc.henneaux@ulb.be}}

\begin{abstract}We analyze the two-dimensional Palatini Gauss-Bonnet theory on an infinite strip (product of a finite interval with the infinite line, corresponding to ``time"). The theory has only boundary degrees of freedom. Its phase space is the cotangent bundle to the group manifold of $SL(2,\mathbf{R})$, subject to a (first-class) constraint quadratic in the momenta.  With the simplest choice of boundary Hamiltonian, namely $H = 0$, the theory is shown to describe geodesics on the group manifold of $SL(2,\mathbf{R})$, with a ``mass" determined by the Palatini Gauss-Bonnet coupling constant. Other choices of boundary Hamiltonians compatible with gauge invariance are also possible.  The symmetry group contains (left and right) group translations on $SL(2,\mathbf{R})$. These  are ``boundary symmetries" from the bulk point of view, one copy acting on one end of the interval, the other copy acting on the other end.  Comments on the quantum theory  are also given.
\end{abstract}

 \maketitle

{\it Introduction}: The complexity of four-dimensional gravity, both at the classical and quantum levels, has motivated the study of lower-dimensional models ($d=3$ and $d=2$). These theories have provided an invaluable testing ground for developing and exploring ideas relevant to four-dimensional gravity. 

Somehow puzzling, the AdS$_2$/CFT$_1$ correspondence remained an open problem for many years.  This picture changed dramatically after the works \cite{Almheiri:2014cka,Maldacena:2016upp,Maldacena:2016hyu} and others, who showed that the breaking of conformal symmetry gives rise to a Schwarzian action describing the boundary dynamics of nearly-AdS$_2$  Jackiw-Teitelboim gravity (JT) \cite{Jackiw:1984je,Teitelboim:1983ux}, and the boundary effective theory obtained after integrating  the microscopic fermionic degrees of freedom in the SYK model. These ideas have been under intense scrutiny is recent years.

A salient feature of JT gravity, as of most $2$-dimensional gravitational models \cite{Grumiller:2002nm}, is the introduction of a scalar field (``dilaton") multiplying the Einstein-Hilbert density, which otherwise defines by itself a trivial theory. 

A different mechanism that also leads to a non-trivial  2-dimensional gravitational action while dispensing with the scalar field was introduced in the recent paper \cite{Banados:2025sww}, where we considered  the ``Palatini Gauss-Bonnet theories". These theories exist in all even spacetime dimensions $D=2n$ and involve as dynamical fields a $GL(2n,\mathbf{R})$-connection ${\Gamma^\alpha}_{\beta\mu}$ and an internal metric $g_{\alpha\beta}$.  

The purpose of this paper is to study in detail the Palatini Gauss-Bonnet theory in  $2$ spacetime dimensions. As JT gravity, this theory has only boundary degrees of freedom (see \cite{Banados:2025sww} for details) and is therefore an interesting holographic laboratory. We shall prove that Palatini Gauss-Bonnet theory formulated on a manifold with two disconnected boundaries (an infinite strip $[r_1,r_2] \times \mathbf{R}$) is equivalent to a one-dimensional relativistic particle propagating on a three dimensional anti-Sitter space. We shall leave for the future a direct comparison with JT gravity and its relation to AdS$_2$/CFT$_1$. This is an important aspect but beyond the scope of this work.

In two dimensions, the action we consider is, 
\begin{equation}
I[g,\Gamma] = k \int  \sqrt{\vert g \vert}\, \epsilon^{\alpha}_{\  \beta} \, R^{\beta}_{\ \alpha}(\Gamma) - \int dt \, H \big\vert^{r_2}_{r_1}  \label{GB2a}
\end{equation}
where   $\epsilon^{\alpha}_{\  \beta}$ is defined by 
$
\epsilon^{\alpha}_{\  \beta} = g^{\alpha \gamma} \epsilon_{\gamma \beta} 
$  
and  $k \not=0$ is a dimensionless coupling constant \cite{Conventions}.  We have allowed for the possibility of a boundary term $- \int dt \, H \big\vert^{r_2}_{r_1}$ at the boundary of the strip. This term will play the role of boundary Hamiltonian (integrated over time) and is necessary to make the variational principle well defined when some fields, in our specific case $\Gamma_0$, do not go to zero at the boundary. Giving $H$   completes the definition of the theory.  We shall come back to its admissible forms below and its connection with the boundary behavior of $\Gamma_0$.

As shown in  \cite{Banados:2025sww}, this action can be rewritten as a constrained $B$-$F$ system \cite{Horowitz:1989ng,Birmingham:1991ty}, which reads, in Hamiltonian form \cite{MertensEtAl}
 \bea
 \hspace{-.5cm} I[B, \Gamma, \Gamma_0, \lambda_1,\lambda_2]  = \int dt \int_{r_1}^{r_2} dr \, {\cal L}  -\int dt H\Big \vert_{r_1}^{r_2} \nonumber\\ 
{\cal L} = \mbox{Tr} \Big(  B \dot \Gamma + \Gamma_0 \nabla B  - \lambda_1 B- \lambda_2 (B^2+ k^2\, \eta \, I_{2\times 2})\Big) \label{actionH}
\eea
where the matrix $B$ belongs to $gl(2)$.  Full equivalence of (\ref{actionH}) with (\ref{GB2a}) was established in \cite{Banados:2025sww}. (We split the 1+1 spacetime connection as  $\Gamma_\mu=\{\Gamma_t, \Gamma\}$ with $\Gamma\equiv\Gamma_r$; $\nabla$ is the radial covariant derivative, $\nabla B \equiv \partial_r B + [\Gamma,B]$; $\eta={|\det{g}| \over \det{g} }=\pm 1$ \cite{Banados:2025sww}.  More on our conventions in \cite{Notations}.)

The kinetic term in the action has the standard $``p \dot{q}"$ canonical form, therefore the spatial connection $\Gamma$ and the $B$-field are canonically conjugate,
\begin{equation}
\{{\Gamma^\alpha}_{\beta1}(r), {B^\gamma}_\delta(r') \} = \delta^\alpha_\delta \delta^\gamma_\beta \delta(r-r') \, .
\end{equation}
The other bulk terms in the action are constraint terms, with respective Lagrange multipliers $\Gamma_0$, $\lambda_1$ and $\lambda_2$.

~

{\it Gauge generators in the presence of boundaries}:  The Gauss constraint, $\nabla B = 0$, involves spatial gradients of the fields leading to ``improper gauge symmetries" \cite{Regge:1974zd,Benguria:1976in} which provide useful insight on the boundary theory. The gauge transformations associated with this constraint are the standard $GL(2)$ gauge transformations, 
\begin{equation}
\delta {\Gamma^\alpha}_{\beta 1} = \nabla_1 {\varepsilon^\alpha}_{\beta } \, , \qquad \delta {B^\alpha}_\beta = {[B, \varepsilon]^\alpha}_\beta.   \label{Eq:GaugeT}
\end{equation}
The canonical generator for these symmetries is 
\begin{equation}
G[\varepsilon] = -\int dr \mbox{Tr}   \left( \varepsilon \nabla B \right) +  \mbox{Tr}   \left( \varepsilon B \right)\Big|^{r_2}_{r_1}. \label{Gb} 
\end{equation}
The boundary terms at $r_1$ and $r_2$ are necessary if $\epsilon(r_1)\neq 0$ (respectively $\epsilon(r_2)\neq 0$) otherwise $G[\varepsilon]$ does not have well-defined variations and the Poisson bracket does not exist \cite{Regge:1974zd}.  

If the matrix $\varepsilon$ is a multiple of the identity at $r_1$  (respectively $r_2$), the boundary term at $r_1$ vanishes because of the linear constraint $\mbox{Tr} (B) = 0$: the transformation is then a proper gauge transformation. Thus, the center $\mathbf{R} I_{2 \times 2}$ defines a proper gauge symmetry.

For arbitrary values of $B$ compatible with the constraints and arbitrary traceless gauge parameters $\epsilon(r_1)$ and $\epsilon(r_2)$, the surface term does not vanish, even when the constraints hold.  The gauge transformations are in that case improper and correspond to transformations having a non trivial physical action \cite{Benguria:1976in}. The improper gauge symmetries are parametrized at each boundary by the quotient algebra, which is $sl(2)$,
\begin{equation}
sl(2) \simeq \frac{gl(2)}{\mathbf{R} I_{2 \times 2}}.
\end{equation}

One might fear that because of the Gauss constraint $\nabla B = 0$, which is a first order equation relating the values of $B(r_2)$ to the value of $B(r_1)$, the boundary terms at the two boundaries are not independent. This would be the case if the group were abelian since then the covariant derivative would coincide with the ordinary derivative and the constraint would imply $B(r_2)=B(r_1)$, leading to the equality of the boundary terms when $\varepsilon(r_1) = \varepsilon(r_2)$.  But in the present case, the relation between $B(r_2)$ and $B(r_1)$ involves the phase space variables so that the boundary terms are functionally independent, as the explicit expressions given below manifestly indicate.

$G[\varepsilon]$ generates gauge transformations on all fields, including itself. This implies 
$\{G[\varepsilon_1],G [\varepsilon_2]\}=\delta_2 G[\varepsilon_1]$ for any  two parameters $\varepsilon_1$ and $\varepsilon_2$. 
One can use this relation for a quick evaluation of the algebra. One finds,
\begin{eqnarray}
\{G[\varepsilon_1],G [\varepsilon_2]\} &=& -\int dr [\varepsilon_1,\varepsilon_2] \nabla B +  [\varepsilon_1,\varepsilon_2] B \Big|^{r_2}_{r_1},  \nonumber\label{GG} \\
&= &G[[\varepsilon_1,\varepsilon_2]],  \label{Eq:BracketG}
\end{eqnarray}  
as expected (Tr symbol omitted). This is an example of the general theorem proved in \cite{Brown:1986ed}, showing that the boundary-term issue arises only once.

Since the values of the gauge parameters at the endpoints are independent, one gets two commuting copies of $sl(2)$, one at each endpoint.  The algebra of improper gauge symmetries is thus $sl(2) \times sl(2)$.  This can be seen explicitly decomposing
the gauge parameter as
\begin{equation}
\varepsilon = \epsilon + \tilde{\epsilon}
\end{equation}
where $\epsilon(r_1) = 0$ and $\tilde{\epsilon}(r_2) = 0$.  The gauge generator decomposes accordingly as $G[\varepsilon] = G[\epsilon] + G[\tilde{\epsilon}]$. One finds from (\ref{Eq:BracketG}) that up to purely bulk terms that vanish with the Gauss constraints, the generators $G[\epsilon]$ and $G[\tilde{\epsilon}]$ have vanishing Poisson brackets and separately fulfill the $sl(2)$ algebra.

The $sl(2)$ canonical charges at each endpoint  are subject to the same quadratic relation $ \mbox{Tr} (B^2+ k^2\, \eta \, I_{2\times 2}) = 0$, which is a restriction on the value of the (quadratic) Casimir in terms of the coupling constant and the signature of the metric $g_{\alpha \beta}$.

~

{\it Boundary action:} One can  reformulate the theory as a $1+0$-dimensional boundary theory by integrating the Gauss constraint ``\`a la Moore-Seiberg"  \cite{Moore:1989yh}.  
Since the Hamiltonian plays no role in the reduction to the boundary theory,  we shall first take $H=0$. This means that the improper gauge transformation that accompanies the time evolution is equal to zero, i.e., that the Lagrange multipliers $\Gamma_0$ vanish at the boundaries,  $\Gamma_0(r_1)=\Gamma_{0}(r_2)=0$. We shall comment later on the case $H \not=0$. 

The trace constraint being algebraic is easily dealt with. It just expresses that the gauge algebra is $sl(2)$.  We can thus use the equivalent description of the system as a $SL(2)$ $BF$ system  and assume from now on, that $\Gamma$ is a $SL(2)$ connection  and  that $B$ is identically traceless.  In that description, the sole constraints are the quadratic constraint and the Gauss constraint.

In order to solve the Gauss constraint, we make the phase space change of variables ($(\Gamma(r), B(r)) \rightarrow (U(r), b(r))$ with $U \in SL(2)$ and $b\in sl(2)$ defined by
\be
\Gamma(r) = U(r)^{-1} \, U(r)'   , \quad
B(r) = U(r)^{-1} b(r) U(r)   \label{bu}
\ee

This is a non-canonical but invertible transformation if we impose $U(r_1) = I_{2\times 2}$, the inverse being given by
\begin{eqnarray}
U(r) =\mbox{P}e^{\int_{r_1}^{r} \Gamma(r) dr} \, , \qquad b(r) =  U(r)  B(r) U(r)^{-1}.
\end{eqnarray}

Under gauge transformations $\Gamma \rightarrow S^{-1} dS + S^{-1} \Gamma S$, the Wilson line $U$ and the new variable $b$ transform as, 
\begin{equation}
\hspace{-.3cm} U(r) \rightarrow S(r_1)^{-1}U(r) S(r) ,\ \, b(r) \rightarrow S(r_1)^{-1} b(r) S(r_1).\label{Wsym}
\end{equation}

In the new variables, the Gauss constraint becomes simply $U(r)^{-1} b'(r) U(r) = 0$, the solution of which is $b = b(t)$. Plugging this solution into the bulk action reduces it to a surface term at $r_2$, which is the searched-for action in $1+0$ dimension,
\begin{equation}
I[V,b; \lambda] = \int dt \, \mbox{Tr} \Big[b \dot V V^{-1} - \lambda (b^2 + k^2 \eta I_{2 \times 2} ) \Big] . \label{IB}
\end{equation}
Here $\lambda(t)$ is 
related to $\lambda_2$ as
$
\lambda(t) := \int_{r_1}^{r_2}  dr\, \lambda_2(r,t) 
$ 
and $V(t) \equiv U(t,r_2)$. To reach (\ref{IB}), we used the relation $\dot \Gamma = \nabla(U^{-1} \dot U )$ that follows from the expression of $\Gamma$, the constraint  $\nabla (U^{-1} b U)=0$  and the boundary condition $U(r_1) = I_{2\times 2} \Rightarrow \dot U(r_1) = 0$.

Varying (\ref{IB}) with respect to $b$, $V$ and $\lambda$ we obtain the equations of motion,
$
\dot V V^{-1}  = 2\lambda b $, $
 \dot b = [\dot V V^{-1}, b]$ and
$ 0 = \mbox{Tr}(b^2 + \eta k^2)$,
respectively. 

All the information extracted via the analysis of charges in bulk functionals can be reobtained from the boundary action. The improper gauge transformations of the original action discussed above appear now as Noether symmetries of the boundary action
 (\ref{IB}), which has indeed an explicit $SL(2)\times SL(2)$ symmetry.
The left $SL(2)$ symmetry (improper gauge transformation at $r_1$, $S(r_1) = h$, $S(r_2) = I_{2 \times 2}$) is given by the transformations
\be
V \rightarrow h^{-1} V \, , \qquad b \rightarrow h^{-1} b h \qquad (h \in SL(2))  \label{Eq:BS1}
\ee
 The right $SL(2)$ symmetry (improper gauge transformation at $r_2$, $S(r_1) =I_{2 \times 2}$, $S(r_2) = g$) reads 
\begin{equation}
V \rightarrow V g \, , \qquad b \rightarrow  b  \qquad (g \in SL(2))  \label{Eq:BS2}
\end{equation}
(with $h$ and $g$ time-independent). Both sets of transformations are
clearly symmetries of (\ref{IB}). 

The corresponding Noether charges coincide exactly with the surface terms associated with the bulk improper gauge symmetries, 
\begin{eqnarray}
Q_\zeta &=& -\mbox{Tr}( \zeta B(r_1)) = -\mbox{Tr}( \zeta b), \\  Q_\epsilon &=& \mbox{Tr}( \zeta B(r_2)) = \mbox{Tr}(\epsilon V^{-1} b V) 
\end{eqnarray}
where we have expanded $h$ and $g$ near the identity as $h = I_{2 \times 2} + \zeta$, $g = I_{2 \times 2} + \epsilon$.  The conservation of these charges, $ {db \over dt}=0$ and $ {d \over dt}(V^{-1}b V) =0 $, can in fact be verified directly from the equations of motion of the boundary theory.

It is tempting to think of the boundary theory as living at $r_2$.  However, the set of asymptotic symmetries defined at the other boundary $r_1$ does act non trivially at $r_2$.  This is because Gauss' law relates the two boundaries, the field $V(r_2)$ itself being non-local.

~

{\it Anti de Sitter formulation:} We now show that the action (\ref{IB}) can be seen as a massive particle propagating on AdS$_3$. 
This is not too surprising given the isomorphism $SL(2) \simeq AdS_3$ \cite{Global}. Indeed, the unit determinant condition of $SL(2)$ matrices,
\be
V = \begin{pmatrix} \alpha & \beta \\ \gamma & \delta \end{pmatrix} \in SL(2) \, , \qquad (\alpha \delta-\beta \gamma = 1) \, ,
\ee
has a one-to-one map with the $\mathbf{R}^{2,2}$ AdS$_3$ embedding $-T_0^2 + X_1^2 + X_2^2 - T_3^2 = - 1$ via $\alpha= T_0 + X_1$, $\delta = T_0 - X_1$, $\beta = T_3 +X_2$, $\gamma = -T_3 + X_2$.  A quick way to see the relation between (\ref{IB}) and a free particle on AdS$_3$ is to note that $b$, and then $\lambda$, can be eliminated from their own equations arriving to the group manifold action $\int dt \sqrt{V^{-1} \dot VV^{-1} \dot V}$. Parametrizing $V$ gives the Lagrangian of a free relativistic particle on AdS$_3$. We choose a route that takes us directly to the Hamiltonian version appropriated to study the quantum theory. 

Let us parametrize $V$ in terms of the familiar AdS global coordinates $\{T,\rho, \Psi\}\equiv q^A$  via the explicit formulas $\alpha = \cosh \rho \cos T + \sinh \rho \cos \Psi$, $
\beta = \cosh \rho \sin T - \sinh \rho \sin \Psi$, $\gamma = -\cosh \rho \sin T - \sinh \rho \sin \Psi$ and $\delta = \cosh \rho \cos T - \sinh \rho \cos \Psi$.

The matrix $b$ lives in $sl(2)$. We parametrize it in terms of three momenta $\{p_0,p_1,p_2\}\equiv \{p_T, p_\rho, p_\Psi\} \equiv \{p_A\}$, 
\begin{eqnarray}
b= {1 \over 2}R P - {1 \over 2}(p_0+p_2) i\sigma_2,
\end{eqnarray}  
($\sigma_2$ is the Pauli matrix) with $R$ and $P$  given by, 
\be
R=\left( \begin{array}{cc}
\cos S & \sin S \\ 
-\sin S & \cos S 
\end{array}  \right)  ,  \quad P = \left(\begin{array}{cc}
p_1 & -P_{12} \\ 
-P_{12} & -p_1
\end{array}  \right).
\ee
Here, $S = T+\Psi$ and $P_{12}\equiv p_0 \tanh\rho + p_2 \coth\rho$.

By direct replacement into the action (\ref{IB}) one finds (with ${1 \over 2}\lambda\rightarrow \lambda$)
\begin{eqnarray}
I = \int dt \Big[ p_A \dot q^A - \lambda \left( \gamma^{AB}p_A p_B +4\eta k^2  \right) \Big],\label{particle}
\end{eqnarray} 
where
\begin{equation}
\gamma_{AB} dq^A dq^A = - \cosh^2(\rho) dT^2 + d\rho^2 + \sinh^2(\rho) d \Psi^2 \label{Eq:MetricAdS}
\end{equation}
is the AdS$_3$ metric \cite{Dimensions}. Thus, as announced, the action (\ref{IB}) is equivalent to a particle propagating on AdS$_3$ with mass-shell condition  $\gamma^{AB}p_A p_B +m^2 = 0$, where the mass is given by
\begin{equation}
m^2 = 4 k^2 \eta \, . \label{mass}
\end{equation}
This is negative for Minkowskian signature of  $g_{\alpha \beta}$ $(\eta=-1)$ and positive for Euclidean signature $(\eta=1)$.  Note that the signature of the two-dimensional metric $g_{\alpha \beta}(x)$ only affects the sign of $m^2$, not the signature of the metric $\gamma_{AB}$, which is always AdS$_3$.

The boundary symmetries identified before are now just the isometries of $AdS_3$, three left translations $L^A_a$ and three right translations $R^A_a$ on the group  \cite{LeftRightTranslations}. Both sets of Killing vectors satisfy the $sl(2)$ algebra in the Lie bracket, 
$
[\xi_a,\xi_b]^A  =\epsilon_{ab}^{\ \ c}\, \xi_c^A$. In the anti-de Sitter parametrization, the boundary/Noether charges read simply
$
Q_a = -\xi^A_a p_A \label{Qxi}
$.

In particular, the $AdS$ energy is the conserved Noether charge associated with the Killing vector $\partial/\partial T$, $E = Q^R_2 + Q^L_2$,  and can be rewritten as 
\begin{equation}
E_{AdS} = -m \boldsymbol{\gamma}\left(\frac{\partial}{\partial T},\frac{dq^M}{d\tau}\right) = -m \gamma_{00}\frac{dT}{d\tau} \, .
\end{equation}
It is positive for future-oriented timelike motions.

The equivalence of the theory with geodesic motion in anti-de Sitter space, and the resulting physical interpretation, naturally leads to the consideration of the covering space $CAdS_3$ as configuration space for the ``particle", in order to avoid closed timelike curves.

~

{\it Non zero Hamiltonian:} Including a non-zero boundary Hamiltonian in  (\ref{GB2a}) is permissible but changes the interpretation of the theory.  The choice of the boundary Hamiltonian goes hand in hand with the choice of the boundary values of the Lagrange multipliers $\Gamma_{0}(r_i)$ associated with the Gauss constraint as most clearly seen in the form (\ref{actionH}) of the action.  Indeed, if $\Gamma_{0}(r_i) \not=0$, one must add a boundary term to cancel the boundary term picked up upon integration by parts in the variation of the action, which is the time integral of $-\Big[\mbox{Tr} (\Gamma_{0}(r) \delta B(r)) \Big]_{r_1}^{r_2}$.  This is necessary in order to have a well-defined action principle \cite{Regge:1974zd}.  Such a boundary term will exist only if $\mbox{Tr} (\Gamma_{0}(r) \delta B(r))$ is integrable at the boundaries, i.e., if $
\Gamma_{0}(r_i) = \frac{\partial H^{(i)}(B_i)}{\partial B_i} \qquad (i=1,2).
$
for some $H^{(i)}(B_i)$ with $B_i \equiv B(r_i)$.  This yields the Hamiltonian 
$H  = H^{(2)}(B_2) - H^{(1)}(B_1)$, or
\be
H = H^{(2)}(Q^R_i) - H^{(1)}(Q^L_i).
\ee

The Hamiltonian $\mathcal H$ is invariant under proper gauge symmetries, as it should.  Because the algebra is not abelian, it is not invariant under improper ones unless one takes $H^{(i)}(B_i) = \mbox{Tr} (B_i^2)$, but given the constraint enforced by $\lambda_2$, this is equivalent to $H^{(i)}(B_i) =$ constant (and $\Gamma_{0}(r_i) = 0$).   Since invariance under improper gauge transformations is not a condition to be imposed on the theory, arbitrary choices of $H^{(i)}(B_i)$ are definitely allowed.  For generic choices of $H$,  one gets in general $\{H, Q^{L,R}_i\}\not=0$ except for the unbroken ``sub-algebra" of transformations that commute with $H$.  The charges associated with the symmetries that do not commute with $H$ will be explicitly time-dependent.    The Hamiltonian being a function of the generators of the algebra, the $sl(2) \times sl(2)$ algebra is nevertherless a symmetry \cite{NewSymmetry}.

~

{\it Quantum theory}:  We follow Dirac prescription for constructing the quantum theory, in which one imposes the constraints on physical states.  In the representation where the group coordinates $q^A$ are diagonal and $\hat p_A=i{\partial \over \partial q^A}$, this constraint on $\phi(q^A)$ becomes the Klein Gordon equation on $AdS_3$,
\begin{equation}
-\Box_q \phi + m^2 \phi = 0.  
\end{equation}
where $\square_q$ is defined with the metric (\ref{Eq:MetricAdS}). Since $\Box_q$ commutes with the AdS generators, the states will transform into representations of the AdS algebra.

If the boundary Hamiltonian $H$ is taken to be zero,  the problem reduces to the much studied question of solving the Klein-Gordon equation in anti-de Sitter space \cite{Fronsdal:1975ac,Avis:1977yn,Wald:1980jn,Breitenlohner:1982jf,Mezincescu:1984ev,Starinets:1998dt}.  
It is then natural to  impose for stability that the $AdS$ energy should be bounded from below.  This means that only the representations from the lowest weight discrete series are allowed (for a review of representations of $SL(2)$ and the $AdS$ groups, see \cite{Bargmann:1946me,Nicolai:1984hb,Kitaev:2017hnr}).
The lowest energy state \cite{Maldacena:1998bw} is 
$
\phi_{h} = c\,  {e^{-2 ih T } \over \cosh^{2h}\rho}$ ($c$ is a normalization constant) and
has energy $E_{AdS}=2h$.  It satisfies $Q^L_-\phi_{h}=Q^R_- \phi_h =0$ (lowering operators) together with $Q^{L_0}\phi_{h}=h\phi_{h}$, $Q^{R_0}\phi_{h} = h \phi_{h}$. Since the eigenvalues of $L_0$ and $R_0$ are equal this state has zero angular momentum (eigenvalue of $Q^{R_0}-Q^{L_0}=i\partial_\Psi$). The wave equation implies the relation,
$
m^2= 4h(h-1)
$
where we recall that the mass is related to the coupling $k$ by (\ref{mass}). 
The descendents have quantized energy levels $E=2h + n$ (with $n$ a positive integer).

The quantum number $h$ must be positive for positivity of the Klein-Gordon norm. From the expression for the mass, we find
$
h = 1 + \sqrt{1+\eta k^2}
$ 
thus, for $\eta=-1$, $k$ must fulfill the Breitenlhoner-Freedman bound \cite{Breitenlohner:1982jf,Mezincescu:1984ev}
\be
k^2 < 1  \, , \quad  (\eta =-1).
\ee
This is the only condition in the case of $CAdS_3$ and $\overline{SL(2)}$. 

If one does not unwrap the closed timelike curves and sticks to the original $AdS_3$, the weight $h$ is restricted by single valuedness  of the wave function  on the time circle by the quantization condition   
$
2h = \mbox{integer} 
$,
that is, $h$ can be integer or semi-integer. This implies that the Palatini Gauss-Bonnet coupling $k$ must also be quantized, since the mass, and hence $k$, is related to $h$ through $
m^2= 4h(h-1)
$. 

If $H \not=0$, the quantum states $\phi(q^A, t)$ also depend on the boundary time $t$ through the Schrödinger equation
$
i \hbar \frac{\partial \phi}{\partial t} = H \phi$ (in addition to the Klein-Gordon equation).
Because  the Hamiltonian does not in general commute with all the generators of $sl(2) \oplus sl(2)$, the energy (defined now as the eigenvalue of $H$) takes  different values within an irreducible representation.  The splitting can be easily computed since $H$ is a function of the $sl(2) \oplus sl(2)$ charges.  

If $H$ is a positive definite function of the AdS charges, such as $H = (E_{AdS})^2$ (say), then it would seem admissible to consider all the solutions of the Klein-Gordon equation, including those corresponding to other (normalizable) representations, such as the continuous principal series, or the discrete highest weight representations (see e.g. \cite{Maldacena:2000hw}).  

To conclude, we have established in this paper the equivalence of the two-dimensional Palatini Gauss-Bonnet theory with the dynamics of a point moving on $AdS_3$ and subject to the Klein-Gordon constraint, with a mass that depends on the signature of the internal metric $g_{\alpha \beta}$ and the coupling constant $k$.  This constraint fully characterizes the system when the boundary Hamiltonian $H$ vanishes, but different choices of $H$ are permissible and define different theories.  Non-vanishing $H$'s remain fully tractable because they depend on the generators of the algebra.

It would be interesting to extend these results to include spin, by starting with a supersymmetric version of the two-dimensional Palatini Gauss-Bonnet.  We leave this for future work.

~

The authors would like to thank the organizers of the NordGrav 2026 meeting at Universidad Arturo Prat, Iquique, Chile, during which an important fraction of this work was done. The work of MH is partially supported by FNRS-Belgium (convention IISN 4.4503.15), as well as by research funds from the Solvay Family. M.B was partially funded by a Grant from Vicerrector\'ia Acad\'emica UC-Chile.


\providecommand{\href}[2]{#2}\begingroup\raggedright


\begin{thebibliography}{10}
\bibitem{Conventions}   $R^{\beta}_{\ \alpha}(\Gamma)$ is the curvature $2$-form of the $GL(2)$ connection ${\Gamma^\alpha}_{\beta \mu}$ and $g$ is the determinant of the internal metric $g_{\alpha \beta}$.  From now on, $GL(2) \equiv GL(2, \mathbf{R})$, $SL(2) \equiv SL(2, \mathbf{R})$ since we deal only with the real linear groups. Furthermore, we take for definiteness the explicit realization of the group elements in terms of $2 \times 2$ real matrices (and not any other representation).
\bibitem{Notations} Our notations are as follows: (i) The matrix $B \equiv {B^\alpha}_\beta {T^\beta}_\alpha$ is the $B$-field transforming as $B \rightarrow S^{-1}BS$ under $GL(2)$ transformations $S$, where ${T^b}_a$ is a basis of generators of the Lie algebra $gl(2)$.  A definite choice is not needed at this stage.  (ii) $\Gamma \equiv {\Gamma^\alpha}_{\beta1} {T^\beta}_\alpha$ is the spatial component of the $GL(2)$-valued connection $1$-form $ {\Gamma^\alpha}_{\beta\mu} {T^\beta}_\alpha dx^\mu$, transforming as $\Gamma \rightarrow S^{-1} dS + S^{-1} \Gamma S$. (iii)  $\Gamma_0 \equiv {\Gamma^\alpha}_{\beta0} {T^\beta}_\alpha$ is its temporal component, which acts as a Lagrange multiplier for the Gauss' constraint $\nabla B = 0$.  (iv) $\nabla B \equiv \nabla_1 B = \partial_1 B + [\Gamma_1, B]$ is the spatial covariant derivative of $B$. (v) $\lambda_1$ and $\lambda_2$ are Lagrange multipliers for the constraints $\mbox{Tr}  B=0$ and  $ \mbox{Tr} (B^2- k^2\, \eta \, I_{2\times 2}) = 0$. (vi) $I_{2\times 2}$ is the identity in 2 dimensions. In the following we shall not systematically write it. (vii) $\eta = \pm 1$ is the sign of the determinant of the metric $g_{\alpha \beta}$.
\bibitem{Global} Note that  $AdS_3$ means here ``wrapped" AdS space (with closed timelike lines) and not its covering space  $CAdS_3$.  The isometry group acts by left and right multiplication as $A \rightarrow LAR$, $A,L, R \in  SL(2)$ and is isomorphic to the quotient $\frac{SL(2) \times SL(2)}{\mathbf{Z}_2}$ where $\mathbf{Z}_2$ is the center of $SL(2)$ (obtained by simultaneously replacing $L$ by $-L$ and $R$ by $-R$).

Since the original action involved only Lie algebra elements, the previous considerations extend straightforwardly to the covering group $\overline{SL(2)}$ and the unwrapped covering space $CAdS_3$.  Going in the opposite direction (taking a quotient rather than unwrapping), one can consider $PSL(2) \simeq \frac{SL(2)}{\mathbf{Z}_2}$, i.e., $AdS_3$ quotientized by the isometry $T_0 \rightarrow - T_0$, $T_3 \rightarrow - T_3$, $X_1 \rightarrow - X_1$, $X_2 \rightarrow - X_2$.  All these groups share the same Lie algebra.
\bibitem{Dimensions} Note that the group coordinates are dimensionless and $AdS$ space can be taken to have unit radius of curvature.
\bibitem{NewSymmetry} In our paper \cite{Banados:2025sww}, we pointed out that the diffeomorphism invariance of the action (\ref{GB2a}) was not an independent symmetry but could be expressed in terms of the gauge symmetries generated by the first class constraints.  Of these, the Gauss constraint generates internal gauge transformations, while the trace constraint generates projective transformations.  The transformation generated by the quadratic constraint was called ``new symmetry" as it was not in the original list of original symmetries. Time diffeomorphisms could be expressed in terms of it, but we were reluctant to identify the new symmetry with time diffeorphisms because this involved dividing by a variable (here, $\lambda$) that was not always different from zero.  

It is interesting to note, however,  that with $H=0$ and the understanding of the dynamics as that of a free particle in $AdS$ space, the quadratic constraint, which is the mass shell condition, does generate reparametrizations in the time $t$.  In that case, identifying the two symmetries is therefore meaningful (the condition $\lambda >0$ on the ``lapse" $\lambda$ makes then sense).   But if the boundary Hamiltonian $H$ is not zero, reparametrizations of the boundary time is not a gauge symmetry and yet, the transformations generated by $\mathcal K$ are proper gauge symmetries.  In fact, translations in $t$ at the boundary are, as we have seen, improper gauge symmetries generated by the Gauss constraint with, in general,  a $B$-dependent parameter, but arbitrary reparametrizations are not symmetries at the boundary.  It follows that there is no hope to express, at the boundary, the proper gauge symmetries generated by $\mathcal K$  in terms of the time diffeomorphisms.
\end{thebibliography}
\end{document}